\documentclass{revtex4-2}
\usepackage{graphicx}

\begin{document}


\title{Short-lived electron excitations in $\rm FeTe_{1-x}Se_x$\ as revealed by microwave absorption}

\author{I.~I.~Gimazov} \email{gimazov@kfti.knc.ru}
\affiliation{Zavoisky Physical-Technical Institute\\ 420029, Kazan, Russian Federation}

\author{N.~M.~Lyadov} 
\affiliation{Zavoisky Physical-Technical Institute\\ 420029, Kazan, Russian Federation}

\author{D.~A.~Chareev} 
\affiliation{Institute of Experimental Mineralogy, Russian Academy of Sciences\\ 142432, Chernogolovka, Moscow District,  Russian Federation}

\author{A.~N.~Vasiliev} 
\affiliation{M.~V. Lomonosov Moscow State University\\ 119991, Moscow, Russian Federation}

\author{Yu.~I.~Talanov} 
\affiliation{Zavoisky Physical-Technical Institute\\ 420029, Kazan, Russian Federation}


\begin{abstract}
{The $\rm Fe_{1+y}Te_{1-x}Se_x$\ single crystals with the various Se/Te ratios were studied by the microwave absorption and direct current resistivity measurements. The comparison of the microwave absorption data and the resistivity versus temperature made it possible to separate  the contributions of two types of spin fluctuations. One of them is due to the anisotropic magnetic (nematic) fluctuations. It is observed over the wide temperature range from $\sim 30$\,K to 150 or 200\,K. In FeSe it has the maximum close to the structural transition temperature. Another MWA anomaly is located in the narrow temperature range above the superconducting transition. It is likely due to the antiferromagnetic fluctuations. Annealing of a sample at the temperature around $300^{\circ}$\,C in the oxygen atmosphere made it possible to exclude the effect of excess iron on the observed anomalies. }
\end{abstract}


\maketitle

\section{Introduction}

FeSe has the simplest chemical formula and structure among the Fe-based superconductors, but its phase diagram is rather reach. A large number of phases with structural, electronic, magnetic and orbital order parameters exist and interplay there (see review \cite{Boehmer18} and references therein). The main problem is to specify what interaction determines the superconducting state formation. 

FeSe transfers into the superconducting state at $T_c \simeq 8.5$\,K at ambient pressure. The structural transition from the tetragonal to orthorhombic structure occurs at ~90\,K (from above) and is accompanied by the establishment of the nematic order \cite{Luo17}. It is characterized by breaking the fourfold rotational symmetry and replacing it with the twofold symmetry of electronic parameters, such as resistivity, magnetic susceptibility {\em etc}. 

The data obtained with NMR \cite{Imai09,Baek14}, neutron scattering \cite{Wang16}, and ARPES \cite{Rhodes18} indicate the effect of spin fluctuations on both the nematic ordering and the superconducting pairing. Though the long-range magnetic order is absent in FeSe at standard conditions, it is easily established by applying pressure \cite{Sun16,Terashima16,Khasanov18}. The strengthening of spin fluctuations \cite{Imai09,Sun17} and the increase in of the superconducting transition temperature (up to 38K) \cite{Terashima16,Sun17} occur simultaneously. In this case, the spin fluctuations have the stripe-shape order implying the nematicity up to ~100K, although the structural transition (orbital ordering) temperature reduces \cite{Wang16}.   

Replacing a part of Se with Te or a part of Fe with other transition elements or intercalation of various chemical compounds into FeSe produce the internal stress in the material (so-called ``chemical pressure''), which acts like the external pressure, in particular it raises the superconducting transition temperature $T_c$. Thus, $\rm FeTe_{1-x}Se_x$\ with $x\approx 0.5$ has $T_c \approx 14$K (see, {\em e.g.} \cite{Liu10}), and doping Cr or V (2\% of Fe) increases $T_c$\ up to \~12\,K \cite{Yadav15}. Intercalating $\rm C_3N_2H_{10}$\ along with Li into FeSe allows the authors of \cite{Sun18} to raise the critical temperature up to 45K at ambient pressure. As for the nematic order, its presence is not so unambiguous in these materials. It can be expected in the form of fluctuations like in FeSe above the structural transition $T=90$K \cite{Luo17}. To detect them it is necessary to use the high-frequency measurement technique. In this work the microwave absorption (MWA) was measured at the frequency of approximately 9.3\,GHz in order to reveal fluctuations of nematic and magnetic order parameters in the $\rm FeTe_{1-x} Se_x$\ crystals with the various Se/Te ratio. Taking into account that the microwave absorption is governed by the ohmic loss in the absence of superconductivity and fluctuations, the comparison of the MWA data and the DC resistivity makes it possible to separate out the fluctuation contributions.

\section{Experimental}

The microwave radiation absorbed by a sample was measured using the standard electron spin resonance (ESR) spectrometer Bruker BER-418s. Microwaves are generated by a clystron and transmitted into the cavity resonator with a sample through a waveguide. The clystron frequency is $\sim 9.2\div 9.5$\,GHz. The spectrometer is sensitive to the weak short-lived electron excitations owing to the lock-in technique of the signal detection and amplification. It requires signal modulation. In order to keep the applied magnetic field to be constant during the MWA measurement, we use the modulation of the incident microwave radiation instead of the applied field modulation built in a conventional ESR spectrometer. The microwave field modulation is realized with the PIN-diode inserted in the waveguide between a clystron and resonator. To perform the temperature variation from a room temperature down to 7\,K a sample is placed inside the helium-gas-flow cryostat tube going through the cavity resonator.

It is known that in the case of a conductive material MWA occurs in the skin-layer, and it is proportional to its volume, as long as the skin-layer thickness $\delta$ is much greater than the mean free path of current carriers $l_e$. This condition is strictly fulfilled in our samples. It will be shown below by calculating $\delta$\ with resistivity data and by comparing it with $l_e$\ known in literature \cite{Pourret11}. Therefore, the variation of the skin-layer thickness with temperature results in the temperature dependence of the MWA signal amplitude $A_{mwa}$\ . Since the skin-layer thickness is determined by the resistivity $\rho$\ as $\delta = c \sqrt{\rho \over 2\pi\omega\mu}$ (here $c$\ is the light speed, $\omega$\ is the frequency, and $\mu$\ is the magnetic permeability). Thus $A_{mwa} \propto \sqrt{\rho}$. This refers to the microwave absorption determined by the ohmic loss. The contributions to MWA due to the fluctuations of various order parameters can be separated by comparing the functions $A_{mwa}(T)$\ and $\sqrt{\rho(T)}$. The resistance of samples is measured using a standard four-probe method at the direct current 3.6\,mA.  

Four {$\rm FeTe_{1-x}Se_x$\ crystals with the various Se/Te ratio were studied. The crystals were grown using the $\rm KCl-AlCl_3$ flux technique. The detailed description of the crystal preparation procedure is presented in Ref.\cite{Chareev13}. The nominal and real composition of the samples under study are presented in Table 1 along with their superconducting transition temperatures. The elemental composition of prepared samples is obtained with the accessory for the energy-dispersive X-ray spectroscopy (EDX) of the scanning electron microscope Carl Zeiss EVO 50 XVP. The EDX analysis revealed excess iron in all samples except for FeSe. (In the general case, the chemical formula has to be written as $\rm Fe_{1+y}Te_{1-x}Se_x$.) The  excess Fe acts in two ways: on the one hand, it leads to the additional current carrier scattering, increases the resistivity and lowers the critical temperature \cite{McQueen09,Sun14}; on the other hand, impurity phases may arise and induce internal stress, which can improve the superconducting properties of a material (as it was shown in Ref.\cite{Hartwig18}). Our $\rm Fe_{1+y}Te_{1-x}Se_x$\ samples have a broad superconducting transition ($3\div5$\,K) with a rather high onset temperature (see Table 1). 

The critical temperature is determined from the magnetic AC susceptibility measurements performed on a home-build susceptometer working at a frequency 1.3\,kHz. According to these data, the superconducting volume fraction is close to 100\% in the FeSe crystal and it decreases considerably as the Te content increases (in compliance with  the muon data \cite{Khasanov09}).

\begin{table}[h]
\caption{Nominal and real composition of the studied samples and their superconducting transition temperatures}
\label{tab1}
\begin{center}
\begin{tabular}{|c|c|c|} \hline
Nominal composition & Real composition & $T_c^{on}$, K \\
\hline 
FeSe & $\rm Fe_{0.994}Se_{1.006}$ & 9.1 \\ \hline
$\rm FeTe_{0.45}Se_{0.55}$ & $\rm Fe_{1.08}Te_{0.44}Se_{0.56}$ & 14.2 \\ \hline
$\rm FeTe_{0.5}Se_{0.5}$ & $\rm Fe_{1.27}Te_{0.54}Se_{0.46}$ & 14.8 \\ \hline
$\rm FeTe_{0.9}Se_{0.1}$ & $\rm Fe_{1.19}Te_{0.91}Se_{0.09}$ & 12.3 \\ \hline
\end{tabular}
\end{center}\
\end{table}

\section{Results and Discussion}

The square root of resistivity $\sqrt{\rho}$\ and MWA versus temperature data obtained on the FeSe crystal are shown in Fig.1. The temperature variation of the resistivity has all features described many times in literature (see, {\em e.\,g.}, \cite{Luo17,Terashima16,Kang16}): the positive slope everywhere over the region above $T_c$\ with the residual resistivity ratio (RRR$=\rho(300{\rm K})/\rho(10{\rm K})$) of about 19; the anomaly near $T_s\simeq 90$\,K due to the tetragonal to orthorhombic structural transition; the sudden jump down to zero at $T_c\simeq 9$\,K. As for the magnitude of $\rho$, it changes from 760\,$\mu\Omega$\,cm at $T=290$\,K down to 40\,$\mu\Omega$\,cm at 10\,K. These correspond to the $\delta$\ decrease from 14\,$\mu$m to 3.3\,$\mu$m. We can state with assurance that these $\delta$\ values are much greater than the mean free path known from literature. In particular, in Ref.\cite{Pourret11} the value of $l_e=3.4$\,nm is adduced for the $\rm FeTe_{0.6}Se_{0.4}$\ with $\rho=350\,\mu\Omega$cm. Even though $l_e$\ is more by 10 times than this in FeSe at low temperature, the relation $\delta\gg l_e$\ is strictly fulfilled. Therefore, the condition of a normal skin-effect is kept over whole temperature range above $T_c$.     

It is known from the literature that applying the pressure of $\sim2$\,GPa results in the long-range antiferromagnetic order at $20\div30$\,K above $T_c$. This is manifested in the resistivity $\rho(T)$\ as an upturn anomaly in the range of $T_c<T<T_N$ \cite{Terashima16,Sun17}. The spin fluctuations intensified by overpressure contribute to the resistivity and enhance it as compared with that at ambient pressure. This effect is observed up to $\sim150$\,K \cite{Sun17}. The fluctuation contribution undistinguished in $\rho(T)$\ at ambient pressure comes out in the high-frequency data, {\em i.\,e.} in the MWA measurements. In Fig.1 the MWA amplitude is attached to the $\rho^{1/2}$\ data by magnitude and slope at high temperatures. Upon lowering the temperature below $\sim 170$\,K the $A_{mwa}(T)$\ curve deflects up from the $\rho^{1/2}(T)$ data owing to the short-lived fluctuations of spin or nematic order. The difference between two functions increases with the temperature decrease and reaches its maximum just below $T_s$. With further lowering the temperature the divergence diminishes and converges on zero near $T=T_c$. The similar behavior was found for the contribution to the in-plane resistivity anisotropy of the FeSe crystal from the nematic order \cite{Tanatar16}. The variation of $\Delta\rho = \rho_a-\rho_b$\ with the temperature decrease was described to be due to oppositely directed impact of the nematic ordering and the inelastic scattering by anisotropic magnetic fluctuations. Whereas the former increases and flattens out at low temperatures, the latter decreases down to zero as $T^2$. 

Close to $T_c$, where the pressure strengthens the spin fluctuations up to the steady antiferromagnetic state, the peak of $A_{mwa}(T)$\ is observed (Fig.1). Note, that the $A_{mwa}(T)$\ curve has the kink at $T=T_s$\ characteristic for the tetragonal to orthorhombic structure transition and sharp fall at $T=T_c$\ the same as $\rho(T)$. It was shown with the inelastic neutron-scattering study of FeSe \cite{Wang16nc} that there are spin fluctuations of two types below $T\sim 180$\,K: anisotropic (stripe type) and isotropic (N\'eel type). They have different temperature dependences. Upon lowering the temperature the N\'eel spin fluctuations reduce while the stripe fluctuations grow. The especially sharp variations of the fluctuation intensity occur in the point of the structural transition $T_s$. Such behavior and competition of two types of spin fluctuations lead apparently to the complex temperature dependence of the fluctuation contribution to MWA with two maximums.
  
\begin{figure}
\begin{center}
\includegraphics[width=82mm]{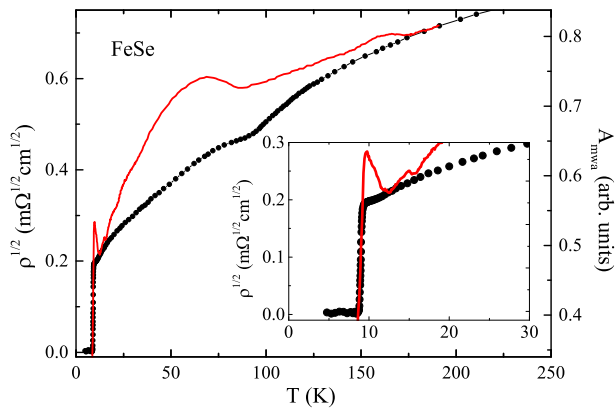}
\end{center}
\caption{Temperature dependence of resistivity in the 1/2 power $\rho^{1/2}(T)$\ (black circles) and the microwave absorption amplitude $A_{mwa}$\ (solid line, red online) for the FeSe crystal. MWA is recorded in the residual field of spectrometer magnet $H_0=25$\,Oe perpendicular to the {\it ab}-plane of crystal. Insert shows the data at the narrow temperature range from 0 to 30\,K}
\label{fig1}
\end{figure}  

It should be pointed out that the peak of $A_{mwa}$\ close to $T_c$\ is observed in cuprate superconductors as well. However, its origin is not the same there. Namely, it is due to superconducting fluctuations in cuprates \cite{Grbic09,Gimazov17}. It was shown earlier, that in iron chalcogenides the superconducting fluctuations exist in a very narrow temperature range (less than 1K) close to  $T_c$\ \cite{Yang17,Nabeshima17}. The MWA peak just above  $T_c$\ has noticeably larger width in temperature (several degrees Kelvin). Moreover, it does not demonstrate the magnetic-field dependence specific for the MWA loss peak due to superconducting fluctuations \cite{Grbic09,Gimazov17}. (The detailed study of the magnetic field effect on MWA in {$\rm FeTe_{1-x}Se_x$\  is the subject of forthcoming article.) Therefore, the connection of the MWA peak in Fig.1 with superconducting fluctuations seems unlikely. It is more probable it is due to the antiferromagnetic fluctuations which induce the long-range ordering under a pressure.    

The partial replacement of selenium with tellurium in the $\rm Fe_{1+y}Te_{1-x}Se_x$\ samples results in following. At first the results obtained for the sample with approximately equal contents of Se and Te are discussed by the example of the $\rm Fe_{1.08}Te_{0.44}Se_{0.56}$\ crystal. The data of the resistivity and MWA measurements for this sample are shown in Fig.2. Since the $\rho$ value is more than that of FeSe, the condition of a normal skin-effect ($\delta\gg l_e$) is fulfilled for $\rm Fe_{1+y}Te_{1-x}Se_x$\ as well. The remarkable difference in the temperature dependence of the resistivity as compared with that of FeSe is clearly seen. First of all, it is of the semiconductor type instead of the metallic one in FeSe. It is an obvious consequence of excess iron in $\rm Fe_{1.08}Te_{0.44}Se_{0.56}$\ \cite{Sun14}. In spite of this, the superconducting transition temperature remains high enough, 14.8\,K. Moreover, there is no kink on the $\rho(T)$\ function signaling about the structural transition and about the establishment of nematic order below $T_s$. Nevertheless, the divergence of two functions, $\rho^{1/2}(T)$\ and $A_{mwa}(T)$\ in the interval of $30{\rm K}<T<140{\rm K}$, indicates that the nematic (anisotropic magnetic) fluctuations affect the microwave absorption in this temperature region. 

\begin{figure}
\begin{center}
\includegraphics[width=82mm]{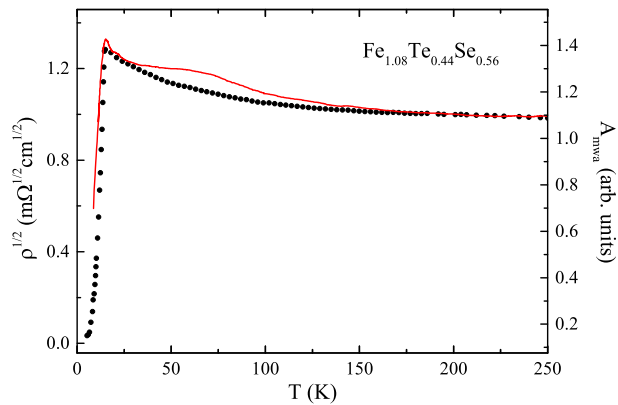}
\end{center}
\caption{Temperature dependence of $\rho^{1/2}(T)$\ (black circles) and  $A_{mwa}$\ (solid line, red online) for the $\rm Fe_{1.08}Te_{0.44}Se_{0.56}$\ crystal ($H_0=25$\,Oe is perpendicular to the {\it ab}-plane of crystal.).}
\label{fig2}
\end{figure}  

Thus it is possible that the internal strain induced by the Se substitution with Te gives rise to the strengthening of spin (nematic) fluctuations, {\em i.\,e.} its impact is similar to that of the external hydrostatic pressure. Also the enhancement of the N\'eel-type fluctuations is manifested through the MWA peak in the narrow temperature range close to the superconducting transition (above it) (see Fig.2). 

Upon further increase in the tellurium fraction up to $\sim 90$\% all effects described above become more pronounced (see Fig.3). The temperature dependence of the resistivity has the same features as that of $\rm Fe_{1.08}Te_{0.44}Se_{0.56}$\ : the semiconducting course of the $\rho(T)$\ function above the superconducting transition temperature and a sharp fall down to zero at  $T_c$. Note, that such pattern of the $\rho(T)$\ dependence is observed for all $\rm Fe_{1+y}Te_{1-x}Se_x$\ samples with $x<0.4$\ \cite{Sales09}. The hump of $A_{mwa}(T)$\ over $\rho^{1/2}(T)$\ has larger height and wider temperature spread (from 30\,K to 200\,K) than those of $\rm Fe_{1.08}Te_{0.44}Se_{0.56}$\ (Fig.2). Since this feature is attributed to the contribution of the anisotropic spin (nematic) fluctuations, it can indicate the strengthening of these fluctuations upon increasing the Te/Se ratio. The narrow peak of  $A_{mwa}$\ near  $T_c$\ can be associated with spin fluctuations of different type.

\begin{figure}
\begin{center}
\includegraphics[width=82mm]{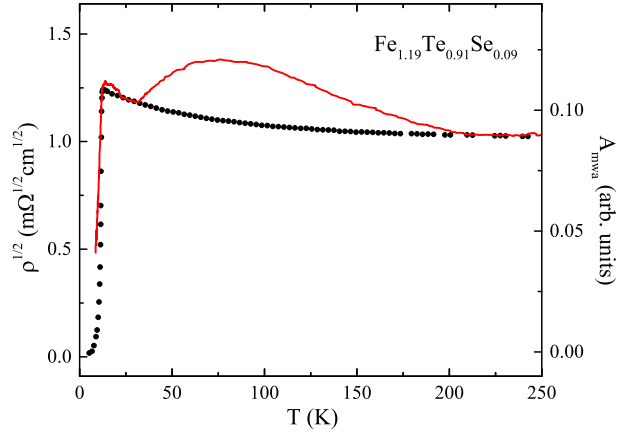}
\end{center}
\caption{Temperature dependence of resistivity $\rho^{1/2}(T)$\ (black circles) and the microwave absorption  $A_{mwa}$\ (solid line, red online) for the $\rm Fe_{1.19}Te_{0.91}Se_{0.09}$\ crystal. (The applied magnetic field $H_0=25$\,Oe is perpendicular to the {\it ab}-plane of crystal.)}
\label{fig3}
\end{figure}  

Our results of the comparative study of the DC resistivity and microwave absorption in the $\rm Fe_{1+y}Te_{1-x}Se_x$\ crystals are in good agreement with the neutron scattering study of spin fluctuations in such samples \cite{Liu10}. The neutron study showed the presence and competition of two types of magnetic correlations ({\em i.\,e.}, fluctuations). First of them, associated with ($\pi$,0) vector in the Brillouin zone, leads to the long-range steady-state antiferromagnetic order upon pressure or upon increasing the Te fraction close to 100\%. They are antagonistic to superconductivity. Another type of dynamic magnetic correlations with a ($\pi$,$\pi$) wave vector are associated with the nearly nesting Fermi surface and thus support the nematic order. It is shown in Ref.\cite{Liu10} that the bulk superconductivity rises just in that area of phase diagram where the ($\pi$,$\pi$) magnetic correlations exist.

It is worth noting the possible effect of the excess iron on the magnetic fluctuations. To clarify this item, excess Fe was removed according to the procedure described in Ref.\cite{Sun14}. This operation consists in annealing a sample at the temperature around $300^{\circ}$\,C with the strictly dosed amount of oxygen. As a result of this procedure performed with the $\rm Fe_{1.27}Te_{0.54}Se_{0.46}$\ the $\rho(T)$\ dependence character has changed from semiconducting to metallic one (see Fig.4a and 4b). In this case, both narrow and broad anomalies of the $A_{mwa}(T)$\ dependence remained unchanged. These observations make it possible to conclude that the excess iron is not the origin of the observed anomalies.

\begin{figure}
\begin{center}
\includegraphics[width=82mm]{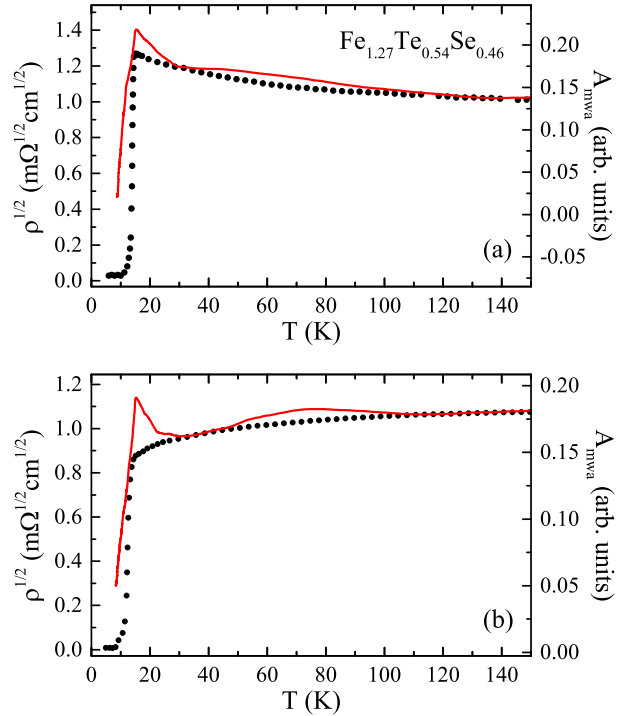}
\end{center}
\caption{Temperature dependence of resistivity $\rho^{1/2}(T)$\ (black circles) and the microwave absorption $A_{mwa}$\ (solid line, red online) for the $\rm Fe_{1.27}Te_{0.54}Se_{0.46}$\ crystal: (a) before annealing in oxygen atmosphere, (b) after annealing.}
\label{fig4}
\end{figure}

In summary, we performed the comparative analysis of the resistivity data and the microwave absorption in single crystals of $\rm Fe_{1+y}Te_{1-x}Se_x$\ iron chalcogenides. The results were obtained in the wide temperature range covering the superconducting transition, the tetragonal to orthorhombic structure transition and the region of magnetic and nematic fluctuations. Upon scaling the $A_{mwa}(T)$\ and $\rho^{1/2}(T)$\ functions at high temperatures, they diverge with lowering a temperature. It is due to the additional contribution from the spin or nematic fluctuations to the microwave absorption which is sensitive to the short-lived electron excitations. Thus, two sectors of the temperature scale with different types of spin fluctuations were located. One of them has the large spread on order of 100\,K. Here the anisotropic magnetic fluctuation of stripe type prevail and induce the nematic order or its fluctuations. The value and the temperature range of the contribution to the MWA from these fluctuations change with the variation of the Te/Se ratio. In pure FeSe the maximum of this contribution is just below the structure transition temperature $T_s$. Another fluctuation area is narrow (several degrees) and located just above the superconducting transition. Here the additional contribution to MWA is supposedly related to the isotropic spin fluctuations. Our findings are in good agreement with the neutron scattering data on spin fluctuations in $\rm Fe_{1+y}Te_{1-x}Se_x$\ \cite{Liu10}. \bigskip

This work was supported by the Russian Academy of Sciences via the grant of Program 1.12 “Fundamental Problems of High-Temperature Superconductivity”.

\bibliographystyle{apsrev4-2}
\bibliography{bibl_rev}

\end{document}